\documentclass[dvipdfmx]{article}
\usepackage{pdfpages}
\usepackage{bm}
\usepackage[left=30mm,right=30mm,top=35mm,bottom=35mm]{geometry}
\usepackage {amsmath}
\usepackage{comment}
\usepackage {float}

\begin{document}
%\begin{comment}
	\begin{center}
		{\bf Three-Dimensional Euler Fluid Code for Fusion Fuel Ignition and Burning}  \\
		\vspace{10truept}
		Hiroki Nakamura, Shigeo Kawata, Ken Uchibori and Takahiro Karino	\\
		\vspace{10truept}
		Utsunomiya University	
		Graduate School of Engineering,  \\
		Yohtoh 7-1-2, Utsunomiya 321-8585, Japan. mc196858@cc.utsunomiya-u.ac.jp, kwt@cc.utsunomiya-u.ac.jp
	\end{center}

{\bf Abstract} \par
	The document describes a numerical algorithm to simulate plasmas and fluids in the 3 dimensional space by the Euler method, in which the spatial meshes are fixed to the space.   The plasmas and fluids move through the spacial Euler mesh boundary. The Euler method can represent a large deformation of the plasmas and fluids. On the other hand, when the plasmas or fluids are compressed to a high density, the spatial resolution should be ensured to describe the density change precisely. The present 3D Euler code is developed to simulate a nuclear fusion fuel ignition and burning. Therefore, the 3D Euler code includes the DT fuel reactions, the alpha particle diffusion, the alpha particle deposition to heat the DT fuel and the DT fuel depletion by the DT reactions, as well as the thermal energy diffusion based on the three-temperature compressible fluid model.
	
\section{Introduction}

In inertial confinement fusion (ICF), the D (deuterium) and T (tritium) fuel should be imploded uniformly to reduce the input driver energy and to release a sufficient fusion energy output. The implosion non-uniformity should be less than a few per cent \cite{kawata1, kawata2}. Recent experimental results demonstrated that the DT fuel is compressed to one thousand to several thousand times the solid density \cite{NIFExperiment1}. The scientific issues include the DT fuel ignition and burning, as well as the implosion uniformity \cite{kawata3}. 

In order to investigate the DT fuel ignition, required are detail experimental \cite{NIFExperiment1}, theoretical \cite{ kawata2} and numerical studies \cite{kawata4, Christiansen}. For the numerical studies, a 2-dimensional code for heavy ion ICF was developed to investigate the implosion non-uniformity smoothing control \cite{ kawata2}. In this document, we present an Euler 3D fluid code algorithm toward a DT fuel ignition and burning. The three-temperature compressible fluid model is employed, together with the DT reaction, the alpha particle generation, diffusion and deposition to sustain the DT reaction in the burning phase \cite{Tahir}. 

%\documentclass{jarticle}
%\begin{document}
\section{Basic Equation}
% \subsection{Equation of continuity}

In this section the basic equations for the compressible plasma are listed below. 

\begin{equation}
\frac{\partial \rho}{\partial t} = - \rho (\nabla \cdot {\bf u}) - ( {\bf u} \cdot \nabla) \rho \label{eq:continuity}
\end{equation}
%\subsection{Equation of motion}
\begin{equation}
\label{eq:motion}
\frac{\partial  {\bf u}}{\partial t} =  -\frac{1}{\rho} \nabla \left( p + q \right) - \left(  {\bf u} \cdot \nabla \right)  {\bf u} 
\end{equation}
%\subsection{Equation of energy}
\begin{eqnarray}
\label{eq:energy1}
\frac{\partial T_i}{\partial t} &=& - \left( {\bf u} \cdot \nabla \right) T_i -\frac{k_B}{C_{V_i}} \left\{ \bigl( \rho B_{T_i} + \frac{p_i + q}{\rho} \bigr) \bigl( \nabla \cdot {\bf u} \bigr) \right\} \\
\label{eq:energy2}
\frac{\partial T_e}{\partial t} &=& - \left( {\bf u} \cdot \nabla \right) T_e -\frac{k_B}{C_{V_e}} \left\{ \bigl( \rho B_{T_e} + \frac{p_e}{\rho} \bigr) \bigl( \nabla \cdot {\bf u} \bigr) \right\} \\
\label{eq:energy3}	
\frac{\partial T_r}{\partial t} &=& - \left( {\bf u} \cdot \nabla \right) T_r -\frac{k_B}{C_{V_r}} \left\{ \bigl( \rho B_{T_r} + \frac{p_r}{\rho} \bigr) \bigl( \nabla \cdot {\bf u} \bigr) \right\}
\end{eqnarray}
Here $\rho$ is the mass density, ${\bf u}=(u, v, w)$ the velocity, $t$ the time, $p_{i,e,r}$ the pressure, $q$ the artificial viscosity, $T_{i, e, r}$ temperature for ion, electron and radiation, $k_B$ is the Boltzmann constant, $C_{V_{i, e, r}}$ is the specific heat for ion, electron and radiation, and $B_{T_{i, e, r}}$ the compressibility for the ion, the electron and the radiation.  

%\end{document} %Basic equation
%\documentclass{jarticle}
%\begin{document}
%\section{Article viscosity}
The artificial viscosities are represented by the following equations.
\begin{eqnarray} \label{eq:viscosity1}
&&q=q_x+q_y+q_z \\ 
&&q_x = \left \{
\begin{array}{l}
 \label{eq:viscosity_x}
\rho C^2_Q \left( \frac{\partial u}{\partial i} \right)^2 + \rho C_L C_s |\frac{\partial u}{\partial i}| \left( \frac{\partial u}{\partial i} < 0 \right) \\
0 \left( \frac{\partial u}{\partial i} \geq 0 \right) \\
\end{array}
\right. \\
 \label{eq:viscosity_y}
&&q_y = \left \{
\begin{array}{l}
\rho C^2_Q \left( \frac{\partial v}{\partial j} \right)^2 + \rho C_L C_s |\frac{\partial v}{\partial j}| \left( \frac{\partial v}{\partial j} < 0 \right) \\
0 \left( \frac{\partial v}{\partial j} \geq 0 \right)
\end{array} 
\right. \\
\label{eq:viscosity_z}
&&q_z = \left \{
\begin{array}{l}
\rho C^2_Q \left( \frac{\partial w}{\partial k} \right)^2 + \rho C_L C_s |\frac{\partial w}{\partial k}| \left( \frac{\partial w}{\partial k} < 0 \right) \\
0 \left( \frac{\partial w}{\partial k} \geq 0 \right)
\end{array}
\right. 
\end{eqnarray}
Here $C_Q=2$, $C_L=1.0$, and $C_s$ is the sound speed in our Euler code. The artificial viscosity $q$ would be a combination of $q_x$ in the $x$ direction, $q_y$ in the $y$ direction and $q_z$ in $z$ direction. 
%\end{document}
 %Article viscosity
%\documentclass{jarticle}
%\begin{document}
\section{Normalization}
The normalization factors of the time $t_0$, the length $L_0$ and the mass $M_0$ are determined as follows:
\begin{eqnarray}
t_0 &=& 1[ns]=10^{-9} [s]\\
L_0 &=& 1[mm] = 10^{-3}[m]\\
M_0&=& \frac{1}{2} (M_D + M_T) = 4.17638 \times 10^{-27} [kg]
\end{eqnarray}
$M_D$ and $M_T$ are the mass of the $D$ atom and the $T$ atom, respectively. Other physical quantities are normalized as follows:
\begin{eqnarray*}
&&L = \tilde{L}L_0 \\
&&t = \tilde{t}t_0 \\
&&j = \tilde{j}j_0 \quad (j_0 = L^3_0)\\
&&u = \tilde{u}u_0 \quad (u_0 = \frac{L_0}{t_0})\\
&&M = \tilde{M}M_0 \\
&&p = \tilde{p}p_0 \quad (p_0 = \frac{M_0}{L_0t^2_0})\\
&&q = \tilde{q}q_0 \quad (q_0 = \frac{M_0}{L_0t^2_0})\\
&&C_v = \tilde{C_v}C_{v_0} \quad (C_{v_0} = \frac{k_B}{M_0})\\
&&B_{T_0} = \tilde{B_T}B_{T_0} \quad (B_{T_0} = \frac{T_0L^3_0}{M^2_0})\\
&&\rho = \tilde{\rho}\rho_0 \quad (\rho_0 = \frac{M_0}{L^3_0})\\
&&T = \tilde{T}T_0 \quad (T_0 = \frac{M_0L^2_0}{t^2_0})\\
\end{eqnarray*}

%\subsection{Normalization of speed}
　The normalization factor of the velocity ${\bf u}=(u, v, w)$ is  $u_0$ = $\frac{L_0}{t_0}$.
\begin{eqnarray}
\tilde{u} = \frac{\partial \tilde{x}}{\partial \tilde{t}}
\end{eqnarray}
\begin{eqnarray}
\tilde{v} = \frac{\partial \tilde{y}}{\partial \tilde{t}} \\
\tilde{w} = \frac{\partial \tilde{z}}{\partial \tilde{t}}
\end{eqnarray}

%\subsection{Normalization of equation of continuity}
The equation of continuity is normalized as follows: 
\begin{equation}
\frac{\partial \rho}{\partial t} = - \rho (\nabla \cdot {\bf u}) - ({\bf u} \cdot \nabla) \rho  \tag{\ref{eq:continuity}}
\end{equation}

\begin{eqnarray}
\frac{\partial \rho}{\partial t} = \rho \left(\frac{\partial u}{\partial x} + \frac{\partial v}{\partial y} + \frac{\partial w}{\partial z}\right) - \left(u \frac{\partial \rho}{\partial x} + v \frac{\partial \rho}{\partial y} + w \frac{\partial \rho}{\partial z}\right)
\end{eqnarray}
Therefore,
\begin{eqnarray}
\footnotesize
\frac{\rho_0}{t_0} \frac{\partial \tilde{\rho}}{\partial \tilde{t}} = - \frac{\rho_0}{L_0} \frac{L_0}{L_0} \frac{L_0}{t_0} \times \tilde{\rho} \left( \frac{\partial \tilde{u}}{\partial \tilde{x}} +\frac{\partial \tilde{v}}{\partial \tilde{y}} + \frac{\partial \tilde{w}}{\partial \tilde{z}} \right) - \frac{L_0}{t_0} \frac{\rho_0}{L_0} \times \left( \tilde{u} \frac{\partial \tilde{\rho}}{\partial \tilde{x}} + \tilde{v} \frac{\partial \tilde{\rho}}{\partial \tilde{y}} + \tilde{w} \frac{\partial \tilde{\rho}}{\partial \tilde{z}}\right). 
\normalsize
\end{eqnarray}
The normalized equation of continuity becomes as follows: 
\begin{eqnarray}
\label{eq:continuity_normal}
\frac{\partial \tilde{\rho}}{\partial \tilde{t}} = - \tilde{\rho} \left( \frac{\partial \tilde{u}}{\partial \tilde{x}} +\frac{\partial \tilde{v}}{\partial \tilde{y}} + \frac{\partial \tilde{w}}{\partial \tilde{z}} \right) - \left( \tilde{u} \frac{\partial \tilde{\rho}}{\partial \tilde{x}} + \tilde{v} \frac{\partial \tilde{\rho}}{\partial \tilde{y}} + \tilde{w} \frac{\partial \tilde{\rho}}{\partial \tilde{z}}\right)
\end{eqnarray}

%\subsection{Normalization of equation of motion and artifical viscosity}
The equation of motion is represented by the following equation.
\begin{equation}
\frac{\partial {\bf u}}{\partial t} = - \left( {\bf u} \cdot \nabla \right) {\bf u} -\frac{1}{\rho} \nabla \left( p + q \right)	\tag{\ref{eq:motion}}
\end{equation}
Each component of the equation of motion is listed here. 
\begin{eqnarray}
\frac{\partial u}{\partial t} &=& -\left( u \frac{\partial u}{\partial x} + v \frac{\partial u}{\partial y} + w \frac{\partial u}{\partial z} \right)  - \frac{1}{\rho} \left( \frac{\partial p_i}{\partial x} + \frac{\partial q}{\partial x} \right) \\
\frac{\partial v}{\partial t} &=& -\left( u \frac{\partial v}{\partial x} + v \frac{\partial v}{\partial y} + w \frac{\partial v}{\partial z} \right)  - \frac{1}{\rho} \left( \frac{\partial p_i}{\partial y} + \frac{\partial q}{\partial y}  \right) \\
\frac{\partial w}{\partial t} &=& -\left( u \frac{\partial w}{\partial x} + v \frac{\partial w}{\partial y} + w \frac{\partial w}{\partial z} \right)  - \frac{1}{\rho} \left( \frac{\partial p_i}{\partial z} + \frac{\partial q}{\partial z} \right)
\end{eqnarray}
These equations are normalized: 
\footnotesize
\begin{eqnarray}
\frac{L_0}{t_0 \cdot T_0} \times \frac{\partial \tilde{u}}{\partial \tilde{t}} &=& - \frac{L_0}{t_0} \frac{L_0}{t_0} \frac{1}{L_0} \times \left( \tilde{u} \frac{\partial \tilde{u}}{\partial \tilde{x}} + \tilde{v} \frac{\partial \tilde{u}}{\partial \tilde{y}} +\tilde{w} \frac{\partial \tilde{u}}{\partial \tilde{z}} \right) - \frac{J_0}{M_0} \times \frac{1}{\tilde{\rho_0}} \left( \frac{p_0}{L_0} \times \frac{\partial \tilde{p_i}}{\partial \tilde{x}} + \frac{q_0}{L_0} \times \frac{\partial \tilde{q}}{\partial \tilde{x}} \right) \\
\frac{L_0}{t_0 \cdot T_0} \times \frac{\partial \tilde{v}}{\partial \tilde{t}} &=& - \frac{L_0}{t_0} \frac{L_0}{t_0} \frac{1}{L_0} \times \left( \tilde{u} \frac{\partial \tilde{v}}{\partial \tilde{x}} + \tilde{v} \frac{\partial \tilde{v}}{\partial \tilde{y}} +\tilde{w} \frac{\partial \tilde{v}}{\partial \tilde{z}} \right) - \frac{J_0}{M_0} \times \frac{1}{\tilde{\rho_0}} \left( \frac{p_0}{L_0} \times \frac{\partial \tilde{p_i}}{\partial \tilde{y}} + \frac{q_0}{L_0} \times \frac{\partial \tilde{q}}{\partial \tilde{y}} \right) \\
\frac{L_0}{t_0 \cdot T_0} \times \frac{\partial \tilde{w}}{\partial \tilde{t}} &=& - \frac{L_0}{t_0} \frac{L_0}{t_0} \frac{1}{L_0} \times \left( \tilde{u} \frac{\partial \tilde{w}}{\partial \tilde{x}} + \tilde{v} \frac{\partial \tilde{w}}{\partial \tilde{y}} +\tilde{w} \frac{\partial \tilde{w}}{\partial \tilde{z}} \right) - \frac{J_0}{M_0} \times \frac{1}{\tilde{\rho_0}} \left( \frac{p_0}{L_0} \times \frac{\partial \tilde{p_i}}{\partial \tilde{z}} + \frac{q_0}{L_0} \times \frac{\partial \tilde{q}}{\partial \tilde{z}} \right) 
\end{eqnarray}
\normalsize
Finally, the equation of motion is normalized as follows: 　
\begin{eqnarray}
\frac{\partial \tilde{u}}{\partial \tilde{t}} &=& -  \left( \tilde{u} \frac{\partial \tilde{u}}{\partial \tilde{x}} + \tilde{v} \frac{\partial \tilde{u}}{\partial \tilde{y}} +\tilde{w} \frac{\partial \tilde{u}}{\partial \tilde{z}} \right) -  \frac{1}{\tilde{\rho}} \left( \frac{\partial \tilde{p} + \partial \tilde{q}}{\partial \tilde{x}}  \right) \\
\frac{\partial \tilde{v}}{\partial \tilde{t}} &=& -  \left( \tilde{u} \frac{\partial \tilde{v}}{\partial \tilde{x}} + \tilde{v} \frac{\partial \tilde{v}}{\partial \tilde{y}} +\tilde{w} \frac{\partial \tilde{v}}{\partial \tilde{z}} \right) -  \frac{1}{\tilde{\rho}} \left( \frac{\partial \tilde{p} + \partial \tilde{q}}{\partial \tilde{y}}  \right) \\
\frac{\partial \tilde{w}}{\partial \tilde{t}} &=& -  \left( \tilde{u} \frac{\partial \tilde{w}}{\partial \tilde{x}} + \tilde{v} \frac{\partial \tilde{w}}{\partial \tilde{y}} +\tilde{w} \frac{\partial \tilde{w}}{\partial \tilde{z}} \right) -  \frac{1}{\tilde{\rho}} \left( \frac{\partial \tilde{p} + \partial \tilde{q}}{\partial \tilde{z}}  \right) 
\end{eqnarray}
\\
The artificial viscosity in 3D is expressed by the following equations. 
\begin{eqnarray}
q_x = \rho C^2_Q \left( \frac{\partial u}{\partial i} \right)^2 + \rho C_L C_s |\frac{\partial u}{\partial i}| \\
q_y = \rho C^2_Q \left( \frac{\partial v}{\partial j} \right)^2 + \rho C_L C_s |\frac{\partial v}{\partial j}| \\
q_z = \rho C^2_Q \left( \frac{\partial w}{\partial k} \right)^2 + \rho C_L C_s |\frac{\partial w}{\partial k}| 
\end{eqnarray}
The artifical viscosity $q_x$ is normalized.
\begin{eqnarray}
\tilde{q_x} = \frac{\rho_0 u^2_0}{q_0} \times \Biggl[ \tilde{\rho} C_Q \left( \frac{\partial \tilde{u}}{\partial i} \right)^2 \Biggr] + \frac{\rho_0 u^2_0}{q_0} \times \biggl[ \tilde{\rho_0} C_L \tilde{C_s} \left| \frac{\partial \tilde{u}}{\partial i} \right| \biggr]
\end{eqnarray}
In addition, 
\begin{eqnarray}
\frac{\rho_0 u^2_0}{q_0} = \frac{ \frac{M_0}{L^3_0} \frac{L^2_0}{t^2_0}}{\frac{M_0}{L_0 t^2_0}} = 1
\end{eqnarray}
Then the normalized artifical viscosities are following: 
\begin{eqnarray}
\tilde{q_x} &=& \tilde{\rho} C^2_Q \left( \frac{\partial \tilde{u}}{\partial i} \right)^2 + \tilde{\rho} C_L \tilde{C_s} \left| \frac{\partial \tilde{u}}{\partial i} \right| \\
\tilde{q_y} &=& \tilde{\rho} C^2_Q \left( \frac{\partial \tilde{v}}{\partial j} \right)^2 + \tilde{\rho} C_L \tilde{C_s} \left| \frac{\partial \tilde{v}}{\partial j} \right| \\
\tilde{q_z} &=& \tilde{\rho} C^2_Q \left( \frac{\partial \tilde{w}}{\partial k} \right)^2 + \tilde{\rho} C_L \tilde{C_s} \left| \frac{\partial \tilde{w}}{\partial k} \right|
\end{eqnarray}

%\subsection{Normalization of equation of energy}
The energy equation is expressed as follows: 
\begin{eqnarray}
\frac{\partial T_i}{\partial t} &=& - \left( {\bf u} \cdot \nabla \right) T_i -\frac{k_B}{C_{V_i}} \left\{ \bigl( \rho B_{T_i} + \frac{p_i + q}{\rho} \bigr) \bigl( \nabla \cdot {\bf u} \bigr) \right\} \\
\frac{\partial T_e}{\partial t} &=& - \left( {\bf u} \cdot \nabla \right) T_e -\frac{k_B}{C_{V_e}} \left\{ \bigl( \rho B_{T_e} + \frac{p_e}{\rho} \bigr) \bigl( \nabla \cdot {\bf u} \bigr) \right\} \\	
\frac{\partial T_r}{\partial t} &=& - \left( {\bf u} \cdot \nabla \right) T_r -\frac{k_B}{C_{V_r}} \left\{ \bigl( \rho B_{T_r} + \frac{p_r}{\rho} \bigr) \bigl( \nabla \cdot {\bf u} \bigr) \right\}
\end{eqnarray}

\begin{comment}
Therefore,
\begin{eqnarray}
\frac{\partial T_i}{\partial t} &=& - \left( u \cdot \nabla \right) T_i -\frac{k_B}{C_{V_i}} \Biggl[ \frac{p_i +q}{\rho} \left( \nabla \cdot u \right)  \Biggr], \\
\frac{\partial T_e}{\partial t} &=& - \left( u \cdot \nabla \right) T_e -\frac{k_B}{C_{V_e}} \Biggl[ \frac{p_e}{\rho} \left( \nabla \cdot u \right)  \Biggr], \\
\frac{\partial T_r}{\partial t} &=& - \left( u \cdot \nabla \right) T_r -\frac{k_B}{C_{V_r}} \Biggl[ \frac{p_r}{\rho} \left( \nabla \cdot u \right)  \Biggr].
\end{eqnarray}
\end{comment}

The ion temperature equation is normalized by the following process, except the term including the compressibility $ B_{T}$. 
\small
\begin{eqnarray}
\frac{T_0}{t_0} \frac{\partial \tilde{T_i}}{\partial \tilde{t}} &=& - \frac{L_0}{t_0} \frac{T_0}{L_0} \left( \tilde{u} \frac{\partial \tilde{T_i}}{\partial \tilde{x}} + \tilde{v} \frac{\partial \tilde{T_i}}{\partial \tilde{y}} + \tilde{w} \frac{\partial \tilde{T_i}}{\partial \tilde{z}} \right) - \frac{M_0}{k_B} \frac{k_B}{\tilde{C_{v_i}}} \Biggl[ \left( \frac{M_0}{L^3_0} \frac{T_0 L^3_0}{M^2_0} \tilde{\rho} \tilde{B_{T_i}} + \frac{M_0}{L_0 t^2_0} \frac{L^3_0}{M_0} \frac{\tilde{p_i}+\tilde{q}}{\tilde{\rho}} \right) \\ \nonumber
&& \times \left\{ \frac{L_0}{t_0} \frac{1}{L_0} \left( \frac{\partial \tilde{u}}{\partial \tilde{x}} + \frac{\partial \tilde{v}}{\partial \tilde{y} } + \frac{\partial \tilde{w}}{\partial \tilde{z}} \right) \right\}  \Biggr]
\end{eqnarray}
\normalsize
The temperature itself is normalized as follows: 
\begin{eqnarray}
T_0 = \frac{M_0 L^2_0}{t^2_0}
\end{eqnarray}
Finally the ion temperature equation is normalized as follows:
\begin{eqnarray}
\frac{\partial \tilde{T_i}}{\partial \tilde{t}} = - \left( \tilde{u} \frac{\partial \tilde{T_i}}{\partial \tilde{x}} + \tilde{v} \frac{\partial \tilde{T_i}}{\partial \tilde{y}} +  \tilde{w} \frac{\partial \tilde{T_i}}{\partial \tilde{z}} \right) - \frac{1}{\tilde{C_{V_i}}} \Biggl[ \left( \tilde{\rho} \tilde{B_{T_i}} +  \frac{\tilde{p_i} + \tilde{q}}{\tilde{\rho}} \right) \left( \frac{\partial \tilde{u}}{\partial \tilde{x}} + \frac{\partial \tilde{v}}{\partial \tilde{y}} + \frac{\partial \tilde{w}}{\partial \tilde{z}} \right) \Biggr]
\end{eqnarray}
Similarly, the equations for the electron temperature and the radiation temperature are normalized as follows: 
\begin{eqnarray}
\frac{\partial \tilde{T_e}}{\partial \tilde{t}} = - \left( \tilde{u} \frac{\partial \tilde{T_e}}{\partial \tilde{x}} + \tilde{v} \frac{\partial \tilde{T_e}}{\partial \tilde{y}} +  \tilde{w} \frac{\partial \tilde{T_e}}{\partial \tilde{z}} \right) - \frac{1}{\tilde{C_{V_e}}} \Biggl[ \left( \tilde{\rho} \tilde{B_{T_e}} + \frac{\tilde{p_e}}{\tilde{\rho}} \right) \left( \frac{\partial \tilde{u}}{\partial \tilde{x}} + \frac{\partial \tilde{v}}{\partial \tilde{y}} + \frac{\partial \tilde{w}}{\partial \tilde{z}} \right) \Biggr] \\
\frac{\partial \tilde{T_r}}{\partial \tilde{t}} = - \left( \tilde{u} \frac{\partial \tilde{T_r}}{\partial \tilde{x}} + \tilde{v} \frac{\partial \tilde{T_r}}{\partial \tilde{y}} +  \tilde{w} \frac{\partial \tilde{T_r}}{\partial \tilde{z}} \right) - \frac{1}{\tilde{C_{V_r}}} \Biggl[ \left(  \tilde{\rho} \tilde{B_{T_r}} +  \frac{\tilde{p_r}}{\tilde{\rho}} \right) \left( \frac{\partial \tilde{u}}{\partial \tilde{x}} + \frac{\partial \tilde{v}}{\partial \tilde{y}} + \frac{\partial \tilde{w}}{\partial \tilde{z}} \right) \Biggr]
\end{eqnarray}
%\end{document}	 %Normalization
\newpage
%\documentclass{jarticle}
%\begin{document}
\section{Discretization} 
\subsection{Discretization of equation of continuity} 
The equation of continuity Eq. (\ref{eq:continuity}) is discretized. 
\begin{equation}
\frac{\partial \rho}{\partial t} = - \rho (\nabla \cdot u) - (u \cdot \nabla) \rho \tag{\ref{eq:continuity}}
\end{equation}
When the equation of continuity is normalized, the following equation is obtained.
\begin{equation}
\frac{\partial \tilde{\rho}}{\partial \tilde{t}} = - \tilde{\rho} \left( \frac{\partial \tilde{u}}{\partial \tilde{x}} +\frac{\partial \tilde{v}}{\partial \tilde{y}} + \frac{\partial \tilde{w}}{\partial \tilde{z}} \right) - \left( \tilde{u} \frac{\partial \tilde{\rho}}{\partial \tilde{x}} + \tilde{v} \frac{\partial \tilde{\rho}}{\partial \tilde{y}} + \tilde{w} \frac{\partial \tilde{\rho}}{\partial \tilde{z}}\right) \tag{\ref{eq:continuity_normal}}
\end{equation}
The left side of this equation is discretized.
\begin{eqnarray}
\left( \frac{\partial \rho}{\partial t} \right)^n_{i+\frac{1}{2},j+\frac{1}{2},k+\frac{1}{2}} = \frac{\rho^{n+1}_{i+\frac{1}{2},j+\frac{1}{2},k+\frac{1}{2}} - \rho^n_{i+\frac{1}{2},j+\frac{1}{2},k+\frac{1}{2}}}{Dt^{n+\frac{1}{2}}_{i+\frac{1}{2},j+\frac{1}{2},k+\frac{1}{2}}}
\end{eqnarray}
Therefore, the discretized equation of continuity is expressed as follows:  
\begin{eqnarray}
\rho^{n+1}_{i+\frac{1}{2},j+\frac{1}{2},k+\frac{1}{2}} &=& \rho^{n}_{i+\frac{1}{2},j+\frac{1}{2},k+\frac{1}{2}} - Dt^{n+\frac{1}{2}} \Biggl[ \rho^n_{i+\frac{1}{2}, j+\frac{1}{2}.k+\frac{1}{2}} \left\{ \left(\frac{\partial u}{\partial x} \right)^n_{i+\frac{1}{2},j+\frac{1}{2},k+\frac{1}{2}} \right. \\ \nonumber
&& \left. + \left(\frac{\partial v}{\partial y} \right)^n_{i+\frac{1}{2},j+\frac{1}{2},k+\frac{1}{2}}+ \left(\frac{\partial w}{\partial z} \right)^n_{i+\frac{1}{2},j+\frac{1}{2},k+\frac{1}{2}} \right\} + \left\{ \left( u \frac{\partial \rho}{\partial x} \right)^n_{i+\frac{1}{2},j+\frac{1}{2},k+\frac{1}{2}} \right. \\ \nonumber
&& \left. + \left(v \frac{\partial \rho}{\partial y} \right)^n_{i+\frac{1}{2},j+\frac{1}{2},k+\frac{1}{2}} +\left( w \frac{\partial \rho}{\partial z} \right)^n_{i+\frac{1}{2},j+\frac{1}{2},k+\frac{1}{2}} \right\} \Biggr] 
\end{eqnarray}
Each term at the right side is shown below.  
\begin{eqnarray*}
&& \left( \frac{\partial u}{\partial x} \right)^n_{i+\frac{1}{2},j+\frac{1}{2},k+\frac{1}{2}} = \frac{ u^n_{i+1,j+\frac{1}{2},k+\frac{1}{2}} - u^n_{i,j+\frac{1}{2},k+\frac{1}{2}}}{D x^n_{i+\frac{1}{2},j+\frac{1}{2},k+\frac{1}{2}}} \\
&& \left( \frac{\partial v}{\partial y} \right)^n_{i+\frac{1}{2},j+\frac{1}{2},k+\frac{1}{2}} = \frac{ v^n_{i+\frac{1}{2},j+1,k+\frac{1}{2}} - u^n_{i+\frac{1}{2},j,k+\frac{1}{2}}}{D y^n_{i+\frac{1}{2},j+\frac{1}{2},k+\frac{1}{2}}} \\
&& \left( \frac{\partial w}{\partial z} \right)^n_{i+\frac{1}{2},j+\frac{1}{2},k+\frac{1}{2}} = \frac{ v^n_{i+\frac{1}{2},j+\frac{1}{2},k+1} - u^n_{i+\frac{1}{2},j+\frac{1}{2},k}}{D z^n_{i+\frac{1}{2},j+\frac{1}{2},k+\frac{1}{2}}} \\
&& \left( u \frac{\partial \rho}{\partial x} \right)^n_{i+\frac{1}{2},j+\frac{1}{2},k+\frac{1}{2}} \\
&& \ \ \ \  = \left\{ 
\begin{array}{ll}
	u^n_{i+\frac{1}{2},j+\frac{1}{2},k+\frac{1}{2}} \frac{\rho^n_{i+\frac{1}{2},j+\frac{1}{2},k+\frac{1}{2}} - \rho^n_{i-\frac{1}{2},j+\frac{1}{2},k+\frac{1}{2}}}{Dx^n_{i,j+\frac{1}{2},k+\frac{1}{2}}} &\left( u^n_{i+\frac{1}{2},j+\frac{1}{2},k+\frac{1}{2}} \geq 0 \right) \\
	u^n_{i+\frac{1}{2},j+\frac{1}{2},k+\frac{1}{2}} \frac{\rho^n_{i+\frac{3}{2},j+\frac{1}{2},k+\frac{1}{2}} - \rho^n_{i+\frac{1}{2},j+\frac{1}{2},k+\frac{1}{2}}}{Dx^n_{i+1, j+\frac{1}{2},k+\frac{1}{2}}} &\left( u^n_{i+\frac{1}{2},j+\frac{1}{2},k+\frac{1}{2}} <0 \right)
\end{array} 
\right. \\
&& \left( v \frac{\partial \rho}{\partial y} \right)^n_{i+\frac{1}{2},j+\frac{1}{2},k+\frac{1}{2}} \\
&& \ \ \ \  = \left\{ 
\begin{array}{ll}
	v^n_{i+\frac{1}{2},j+\frac{1}{2},k+\frac{1}{2}} \frac{\rho^n_{i+\frac{1}{2},j+\frac{1}{2},k+\frac{1}{2}} - \rho^n_{i+\frac{1}{2},j-\frac{1}{2},k+\frac{1}{2}}}{Dy^n_{i+\frac{1}{2},j,k+\frac{1}{2}}} &\left( v^n_{i+\frac{1}{2},j+\frac{1}{2},k+\frac{1}{2}} \geq 0 \right) \\
	v^n_{i+\frac{1}{2},j+\frac{1}{2},k+\frac{1}{2}} \frac{\rho^n_{i+\frac{3}{2},j+\frac{1}{2},k+\frac{1}{2}} - \rho^n_{i+\frac{1}{2},j+\frac{1}{2},k+\frac{1}{2}}}{Dy^n_{i+\frac{1}{2}, j+1,k+\frac{1}{2}}} &\left( v^n_{i+\frac{1}{2},j+\frac{1}{2},k+\frac{1}{2}} <0 \right)
\end{array} 
\right. \\
&& \left( w \frac{\partial \rho}{\partial z} \right)^n_{i+\frac{1}{2},j+\frac{1}{2},k+\frac{1}{2}} \\
&& \ \ \ \  = \left\{ 
\begin{array}{ll}
	w^n_{i+\frac{1}{2},j+\frac{1}{2},k+\frac{1}{2}} \frac{\rho^n_{i+\frac{1}{2},j+\frac{1}{2},k+\frac{1}{2}} - \rho^n_{i+\frac{1}{2},j+\frac{1}{2},k-\frac{1}{2}}}{Dz^n_{i+\frac{1}{2},j+\frac{1}{2},k}} &\left( w^n_{i+\frac{1}{2},j+\frac{1}{2},k+\frac{1}{2}} \geq 0 \right) \\
	w^n_{i+\frac{1}{2},j+\frac{1}{2},k+\frac{1}{2}} \frac{\rho^n_{i+\frac{1}{2},j+\frac{1}{2},k+\frac{3}{2}} - \rho^n_{i+\frac{1}{2},j+\frac{1}{2},k+\frac{1}{2}}}{Dz^n_{i+\frac{1}{2},j+\frac{1}{2},k+1}} &\left( w^n_{i+\frac{1}{2},j+\frac{1}{2},k+\frac{1}{2}} < 0 \right) \\
\end{array} 
\right.
\end{eqnarray*}
\begin{figure}[H]
	\centering
	\includegraphics[width=8.cm]{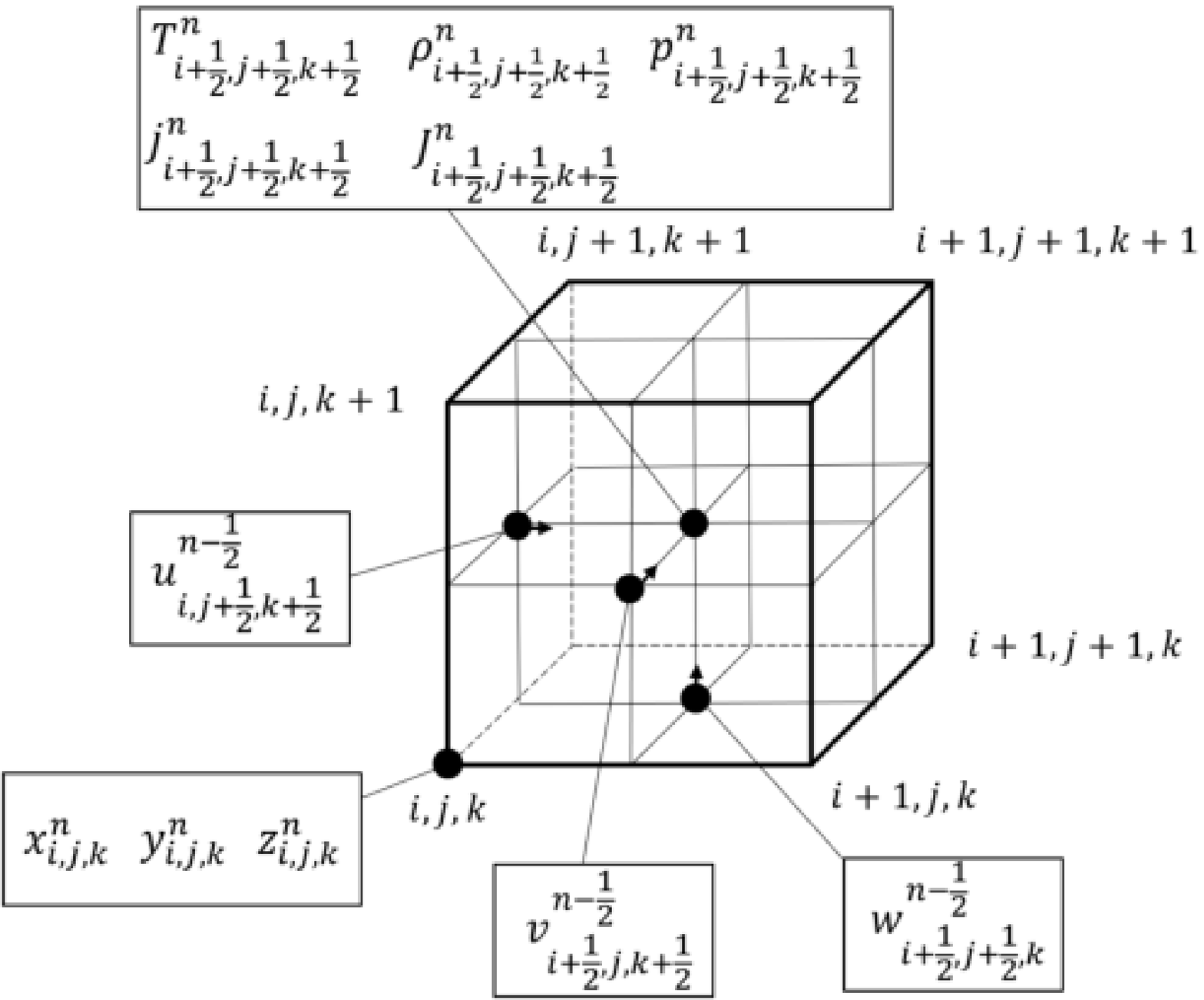}\\
	\ \\
	\includegraphics[width=8.cm]{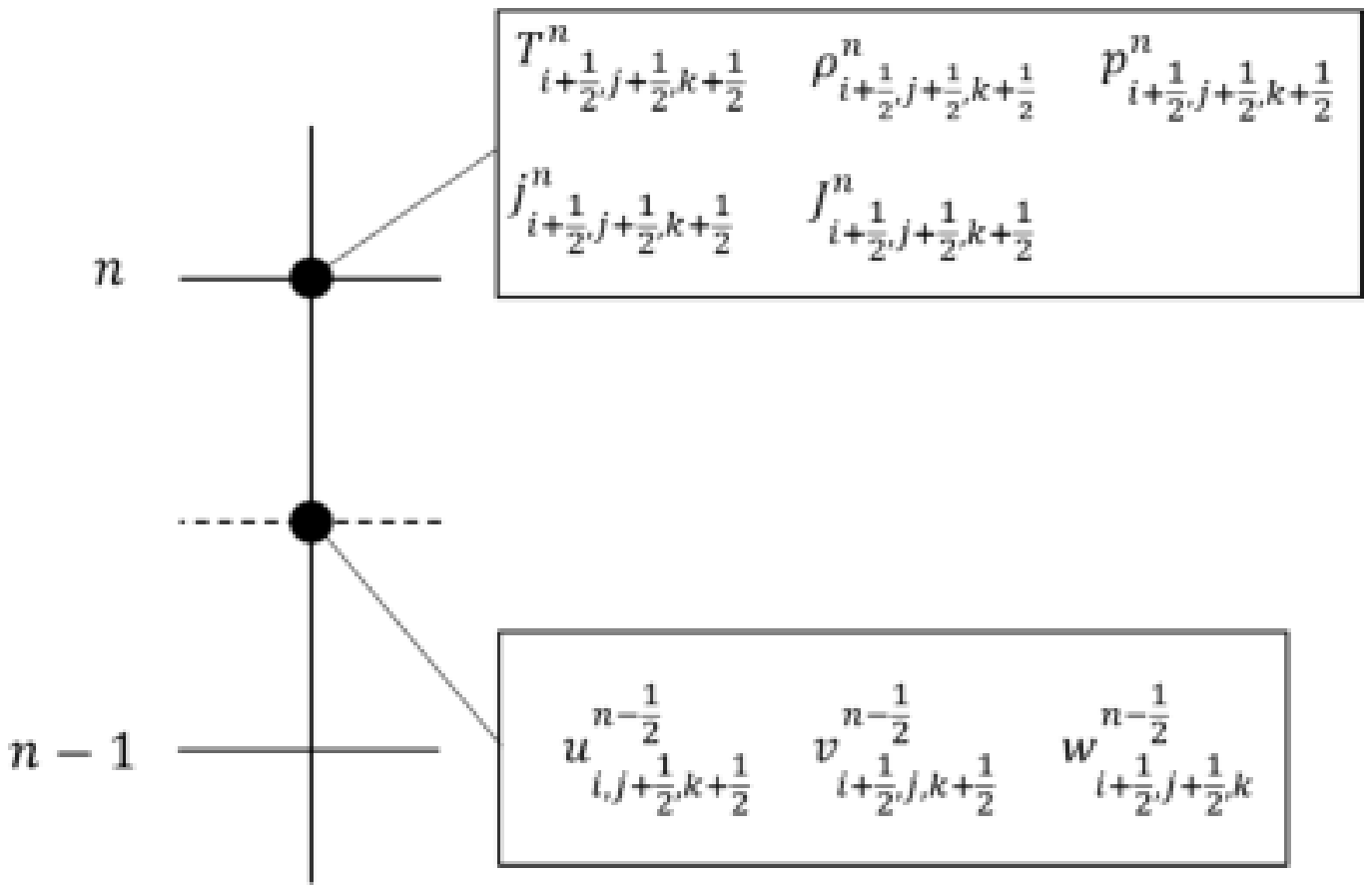}
	\caption{Definition position of each physical quantity}
\end{figure}
\subsection{Discretization of equation of motion} 
The basic equation of the equation of motion is expressed by the following equation.
\begin{equation}
\frac{\partial u}{\partial t} =  -\frac{1}{\rho} \nabla \left( p + q \right) - \left( u \cdot \nabla \right) u \tag{\ref{eq:motion}}
\end{equation}
When this equation is normalized, the following equation is obtained: 
\begin{eqnarray}
\frac{\partial \tilde{u}}{\partial \tilde{t}} &=& -  \left( \tilde{u} \frac{\partial \tilde{u}}{\partial \tilde{x}} + \tilde{v} \frac{\partial \tilde{u}}{\partial \tilde{y}} +\tilde{w} \frac{\partial \tilde{u}}{\partial \tilde{z}} \right) -  \frac{1}{\tilde{\rho}}  \frac{\partial \left(\tilde{p} + \tilde{q}\right)}{\partial \tilde{x}}   \\
\frac{\partial \tilde{v}}{\partial \tilde{t}} &=& -  \left( \tilde{u} \frac{\partial \tilde{v}}{\partial \tilde{x}} + \tilde{v} \frac{\partial \tilde{v}}{\partial \tilde{y}} +\tilde{w} \frac{\partial \tilde{v}}{\partial \tilde{z}} \right) -  \frac{1}{\tilde{\rho}} \frac{\partial \left(\tilde{p} + \tilde{q}\right)}{\partial \tilde{y}}   \\
\frac{\partial \tilde{w}}{\partial \tilde{t}} &=& -  \left( \tilde{u} \frac{\partial \tilde{w}}{\partial \tilde{x}} + \tilde{v} \frac{\partial \tilde{w}}{\partial \tilde{y}} +\tilde{w} \frac{\partial \tilde{w}}{\partial \tilde{z}} \right) -  \frac{1}{\tilde{\rho}} \ \frac{\partial \left(\tilde{p} +  \tilde{q}\right)}{\partial \tilde{z}}   
\end{eqnarray}
The time derivative of $u$, that is, the $x$ component of the velocity becomes as follows: 
\begin{eqnarray}
\left( \frac{\partial u}{\partial t} \right)^n_{i+\frac{1}{2},j+\frac{1}{2},k+\frac{1}{2}} = \frac{u^{n+1}_{i+\frac{1}{2},j\frac{1}{2},k\frac{1}{2}} - u^n_{i+\frac{1}{2},j\frac{1}{2},k\frac{1}{2}}}{Dt^{n+\frac{1}{2}}}
\end{eqnarray}
Therefore, the discretized equation of motion is as follows: 
\begin{eqnarray}
u^{n+\frac{1}{2}}_{i,j+\frac{1}{2},k+\frac{1}{2}} &=& u^{n-\frac{1}{2}}_{i,j+\frac{1}{2},k+\frac{1}{2}} - Dt^n \Biggl[ \left\{ \left(u \frac{\partial u}{\partial x} \right)^n_{i,j+\frac{1}{2},k+\frac{1}{2}} + \left(v \frac{\partial u}{\partial y} \right)^n_{i,j+\frac{1}{2},k+\frac{1}{2}} \right. \\  \nonumber
&& \left.  + \left(w \frac{\partial u}{\partial z} \right)^n_{i,j+\frac{1}{2},k+\frac{1}{2}} \right\} + \left\{ \frac{1}{\rho} \frac{\partial (p+q)}{\partial x} \right\}^n_{i,j+\frac{1}{2},k+\frac{1}{2}} \Biggr] 
\end{eqnarray}

\begin{eqnarray*}
&& \left( u \frac{\partial u}{\partial x} \right)^n_{i,j+\frac{1}{2},k+\frac{1}{2}} \\
&& \ \ \ \ \ \  = \left\{ 
\begin{array}{ll}
u^{n-\frac{1}{2}}_{i,j+\frac{1}{2},k+\frac{1}{2}} \frac{u^{n-\frac{1}{2}}_{i,j+\frac{1}{2},k+\frac{1}{2}} - u^{n-\frac{1}{2}}_{i-1,j+\frac{1}{2},k+\frac{1}{2}}}{Dx^n_{i-\frac{1}{2},j+\frac{1}{2},k+\frac{1}{2}}} &\left( u^{n-\frac{1}{2}}_{i,j+\frac{1}{2},k+\frac{1}{2}} \geq 0 \right) \\
	u^{n-\frac{1}{2}}_{i,j+\frac{1}{2},k+\frac{1}{2}} \frac{u^{n-\frac{1}{2}}_{i+1,j+\frac{1}{2},k+\frac{1}{2}} - u^{n-\frac{1}{2}}_{i+1,j+\frac{1}{2},k+\frac{1}{2}}}{Dx^n_{i-\frac{1}{2},j+\frac{1}{2},k+\frac{1}{2}}} &\left( u^{n-\frac{1}{2}}_{i,j+\frac{1}{2},k+\frac{1}{2}} <0 \right)
\end{array} 
\right. \\
&& \left( v \frac{\partial u}{\partial y} \right)^n_{i,j+\frac{1}{2},k+\frac{1}{2}} \\
&& \ \ \ \ \ \  = \left\{ 
\begin{array}{ll}
	v^{n-\frac{1}{2}}_{i,j+\frac{1}{2},k+\frac{1}{2}} \frac{u^{n-\frac{1}{2}}_{i,j+\frac{1}{2},k+\frac{1}{2}} - u^{n-\frac{1}{2}}_{i,j-\frac{1}{2},k+\frac{1}{2}}}{Dy^n_{i-\frac{1}{2},j+\frac{1}{2},k+\frac{1}{2}}} &\left( v^{n-\frac{1}{2}}_{i,j+\frac{1}{2},k+\frac{1}{2}} \geq 0 \right) \\
	v^{n-\frac{1}{2}}_{i,j+\frac{1}{2},k+\frac{1}{2}} \frac{u^{n-\frac{1}{2}}_{i,j+\frac{3}{2},k+\frac{1}{2}} + u^{n-\frac{1}{2}}_{i,j+\frac{1}{2},k+\frac{1}{2}}}{Dy^n_{i-\frac{1}{2},j+\frac{1}{2},k+\frac{1}{2}}} &\left( v^{n-\frac{1}{2}}_{i,j+\frac{1}{2},k+\frac{1}{2}} < 0 \right) 
\end{array} 
\right. \\
&& \left( w \frac{\partial u}{\partial z} \right)^n_{i,j+\frac{1}{2},k+\frac{1}{2}} \\
&& \ \ \ \ \ \  = \left\{ 
\begin{array}{ll}
	w^{n-\frac{1}{2}}_{i,j+\frac{1}{2},k+\frac{1}{2}} \frac{u^{n-\frac{1}{2}}_{i,j+\frac{1}{2},k+\frac{1}{2}} - u^{n-\frac{1}{2}}_{i,j+\frac{1}{2},k-\frac{1}{2}}}{Dz^n_{i-\frac{1}{2},j+\frac{1}{2},k+\frac{1}{2}}} &\left( w^{n-\frac{1}{2}}_{i,j+\frac{1}{2},k+\frac{1}{2}} \geq 0 \right) \\
	w^{n-\frac{1}{2}}_{i,j+\frac{1}{2},k+\frac{1}{2}} \frac{u^{n-\frac{1}{2}}_{i,j+\frac{1}{2},k+\frac{3}{2}} - u^{n-\frac{1}{2}}_{i,j+\frac{1}{2},k+\frac{1}{2}}}{Dz^n_{i-\frac{1}{2},j+\frac{1}{2},k+\frac{1}{2}}} &\left( w^{n-\frac{1}{2}}_{i,j+\frac{1}{2},k+\frac{1}{2}} < 0 \right) 
\end{array} 
\right. 
\end{eqnarray*}
\begin{eqnarray*}
\left\{ \frac{1}{\rho} \frac{\partial(p+q)}{\partial x} \right\}^n_{i,j+\frac{1}{2},k+\frac{1}{2}} &=& \frac{2}{\rho^n_{i+\frac{1}{2},j+\frac{1}{2},k+\frac{1}{2}} + \rho^n_{i-\frac{1}{2},j+\frac{1}{2},k+\frac{1}{2}}}  \\
&& \left\{ \frac{p^n_{i+\frac{1}{2},j+\frac{1}{2},k+\frac{1}{2}} + q^n_{i+\frac{1}{2},j+\frac{1}{2},k+\frac{1}{2}} - \left(p^n_{i-\frac{1}{2},j+\frac{1}{2},k+\frac{1}{2}} + q^n_{i-\frac{1}{2},j+\frac{1}{2},k+\frac{1}{2}} \right) }{Dx_{i,j+\frac{1}{2},k+\frac{1}{2}}} \right\}
\end{eqnarray*}
The $y$ direction and the $z$ direction are similarly represented.
\\
\begin{figure}[H]
	\centering
	\includegraphics[width=11.cm]{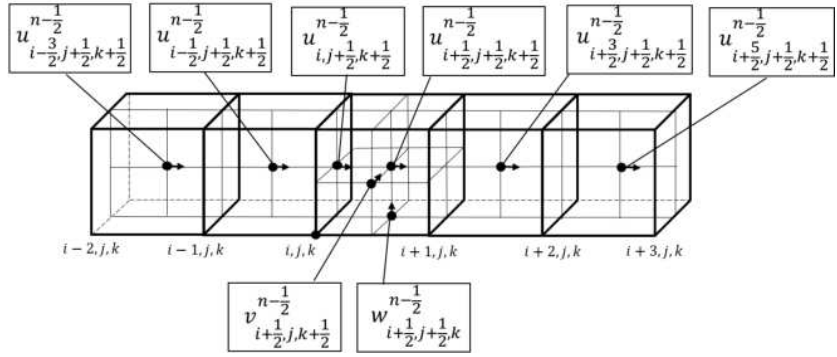}
	\caption{Definition position of velosity in $x$ direction}
\end{figure}

\subsection{Discretization of artificial viscosity}
The artificial viscosity is represented by the following equation.
\begin{eqnarray}\label{eq:viscosity1}
&&q=q_x+q_y+q_z \\
&&q_x = \left \{
\begin{array}{l}
\label{eq:viscosity_x}
\rho C^2_Q \left( \frac{\partial u}{\partial i} \right)^2 + \rho C_L C_s |\frac{\partial u}{\partial i}| \left( \frac{\partial u}{\partial i} < 0 \right) \\
0 \left( \frac{\partial u}{\partial i} \geq 0 \right) \\
\end{array}
\right. \\
&&q_y = \left \{
\begin{array}{l}
\label{eq:viscosity_y}
\rho C^2_Q \left( \frac{\partial v}{\partial j} \right)^2 + \rho C_L C_s |\frac{\partial v}{\partial j}| \left( \frac{\partial v}{\partial j} < 0 \right) \\
0 \left( \frac{\partial v}{\partial j} \geq 0 \right)
\end{array}
\right. \\
&&q_z = \left \{
\begin{array}{l}
\label{eq:viscosity_z}
\rho C^2_Q \left( \frac{\partial w}{\partial k} \right)^2 + \rho C_L C_s |\frac{\partial w}{\partial k}| \left( \frac{\partial w}{\partial k} < 0 \right) \\
0 \left( \frac{\partial w}{\partial k} \geq 0 \right)
\end{array}
\right. 
\end{eqnarray}
Here, $C_Q=2$, $C_L=1.0$, $C_s$ is the sound speed．
When these equations are standardized, the following equations are obtained.
\begin{eqnarray}
\tilde{q_x} &=& \tilde{\rho} C^2_Q \left( \frac{\partial \tilde{u}}{\partial i} \right)^2 + \tilde{\rho} C_L \tilde{C_s} \left| \frac{\partial \tilde{u}}{\partial i} \right| \\ 
\tilde{q_y} &=& \tilde{\rho} C^2_Q \left( \frac{\partial \tilde{v}}{\partial j} \right)^2 + \tilde{\rho} C_L \tilde{C_s} \left| \frac{\partial \tilde{v}}{\partial j} \right| \\ 
\tilde{q_z} &=& \tilde{\rho} C^2_Q \left( \frac{\partial \tilde{w}}{\partial k} \right)^2 + \tilde{\rho} C_L \tilde{C_s} \left| \frac{\partial \tilde{w}}{\partial k} \right| 
\end{eqnarray}
When these equations are discretized, they are expressed by the following equations.
\begin{eqnarray}
{q_x}^n_{i+\frac{1}{2},j+\frac{1}{2},k+\frac{1}{2}} &=& \rho^n_{i+\frac{1}{2},j+\frac{1}{2},k+\frac{1}{2}} C_Q^2 (u^n_{i+1,j+\frac{1}{2},k+\frac{1}{2}} - u^n_{i,j+\frac{1}{2},k+\frac{1}{2}})^2 \\\nonumber
&& + \rho^n_{i+\frac{1}{2},j+\frac{1}{2},k+\frac{1}{2}} C_L C_s \left| u^n_{i+1,j+\frac{1}{2},k+\frac{1}{2}}-u^n_{i,j+\frac{1}{2},k+\frac{1}{2}} \right| \\
{q_y}^n_{i+\frac{1}{2},j+\frac{1}{2},k+\frac{1}{2}} &=& \rho^n_{i+\frac{1}{2},j+\frac{1}{2},k+\frac{1}{2}} C_Q^2 (u^n_{i+\frac{1}{2},j+1,k+\frac{1}{2}} - u^n_{i+\frac{1}{2},j,k+\frac{1}{2}})^2 \\ \nonumber
&& + \rho^n_{i+\frac{1}{2},j+\frac{1}{2},k+\frac{1}{2}} C_L C_s \left| u^n_{i+\frac{1}{2},j+1,k+\frac{1}{2}}-u^n_{i+\frac{1}{2},j,k+\frac{1}{2}} \right| \\
{q_z}^n_{i+\frac{1}{2},j+\frac{1}{2},k+\frac{1}{2}} &=& \rho^n_{i+\frac{1}{2},j+\frac{1}{2},k+\frac{1}{2}} C_Q^2 (u^n_{i+\frac{1}{2},j+\frac{1}{2},k+1} - u^n_{i+\frac{1}{2},j+\frac{1}{2},k})^2 \\ \nonumber
&& + \rho^n_{i+\frac{1}{2},j+\frac{1}{2},k+\frac{1}{2}} C_L C_s \left| u^n_{i+\frac{1}{2},j+\frac{1}{2},k+1}-u^n_{i+\frac{1}{2},j+\frac{1}{2},k} \right| \\
\end{eqnarray}
\\
\subsection{Discretization of energy equation}\label{Discretization of energy equation}
The basic equation of the energy equation is represented by the following equation.
\begin{eqnarray}
\frac{\partial T_i}{\partial t} &=& - \left( u \cdot \nabla \right) T_i -\frac{k_B}{C_{V_i}} \Biggl[ \left( \rho B_{T_i} + \frac{p_i +q}{\rho} \right) \left( \nabla \cdot u \right)  \Biggr] \\
\frac{\partial T_e}{\partial t} &=& - \left( u \cdot \nabla \right) T_e -\frac{k_B}{C_{V_e}} \Biggl[ \left( \rho B_{T_e} + \frac{p_e}{\rho} \right) \left( \nabla \cdot u \right)  \Biggr] \\
\frac{\partial T_r}{\partial t} &=& - \left( u \cdot \nabla \right) T_r -\frac{k_B}{C_{V_r}} \Biggl[ \left( \rho B_{T_r} + \frac{p_r}{\rho} \right) \left( \nabla \cdot u \right)  \Biggr]
\end{eqnarray}
When these equations are normalized, they are expressed by the following equations.
\begin{eqnarray}
\frac{\partial \tilde{T_i}}{\partial \tilde{t}} &=& - \left( \tilde{u} \frac{\partial \tilde{T_i}}{\partial \tilde{x}} + \tilde{v} \frac{\partial \tilde{T_i}}{\partial \tilde{y}} +  \tilde{w} \frac{\partial \tilde{T_i}}{\partial \tilde{z}} \right) - \frac{1}{\tilde{C_{V_i}}} \Biggl[ \left( \frac{\tilde{p_i} + \tilde{q}}{\tilde{\rho}} \right) \left( \frac{\partial \tilde{u}}{\partial \tilde{x}} + \frac{\partial \tilde{v}}{\partial \tilde{y}} + \frac{\partial \tilde{w}}{\partial \tilde{z}} \right) \Biggr] \\
\frac{\partial \tilde{T_e}}{\partial \tilde{t}} &=& - \left( \tilde{u} \frac{\partial \tilde{T_e}}{\partial \tilde{x}} + \tilde{v} \frac{\partial \tilde{T_e}}{\partial \tilde{y}} +  \tilde{w} \frac{\partial \tilde{T_e}}{\partial \tilde{z}} \right) - \frac{1}{\tilde{C_{V_e}}} \Biggl[ \left(  \tilde{\rho} \tilde{B_{T_e}} + \frac{\tilde{p_e}}{\tilde{\rho}} \right) \left( \frac{\partial \tilde{u}}{\partial \tilde{x}} + \frac{\partial \tilde{v}}{\partial \tilde{y}} + \frac{\partial \tilde{w}}{\partial \tilde{z}} \right) \Biggr] \\
\frac{\partial \tilde{T_r}}{\partial \tilde{t}} &=& - \left( \tilde{u} \frac{\partial \tilde{T_r}}{\partial \tilde{x}} + \tilde{v} \frac{\partial \tilde{T_r}}{\partial \tilde{y}} +  \tilde{w} \frac{\partial \tilde{T_r}}{\partial \tilde{z}} \right) - \frac{1}{\tilde{C_{V_r}}} \Biggl[ \left(  \tilde{\rho} \tilde{B_{T_r}} + \frac{\tilde{p_r}}{\tilde{\rho}} \right) \left( \frac{\partial \tilde{u}}{\partial \tilde{x}} + \frac{\partial \tilde{v}}{\partial \tilde{y}} + \frac{\partial \tilde{w}}{\partial \tilde{z}} \right) \Biggr]
\end{eqnarray}
Here, $B_{T_i} = 0$. The ion temperature equation is expressed by the following equation when the left side is discretized.
\begin{eqnarray}
\left( \frac{\partial T_i}{\partial x} \right)^n_{i+\frac{1}{2},j+\frac{1}{2},k+\frac{1}{2}} = \frac{{T_i}^{n+1}_{i+\frac{1}{2},j+\frac{1}{2},k+\frac{1}{2}} - {T_i}^n_{i+\frac{1}{2},j+\frac{1}{2},k+\frac{1}{2}}}{Dt^{n+\frac{1}{2}}}
\end{eqnarray}
Therefore, the energy equation for the discretized ion temperature is expressed by the following equation.
\begin{eqnarray}
{T_i}^{n+1}_{i+\frac{1}{2},j+\frac{1}{2},k+\frac{1}{2}} &=& {T_i}^n_{i+\frac{1}{2},j+\frac{1}{2},k+\frac{1}{2}} - Dt^{n+\frac{1}{2}} \Biggl[ \left\{ \left( u \frac{\partial T_i}{\partial x} \right)^n_{i+\frac{1}{2},j+\frac{1}{2},k+\frac{1}{2}} \right. \\ \nonumber
&& \left. + \left( v \frac{\partial T_i}{\partial y} \right)^n_{i+\frac{1}{2},j+\frac{1}{2},k+\frac{1}{2}} + \left( w \frac{\partial T_i}{\partial z} \right)^n_{i+\frac{1}{2},j+\frac{1}{2},k+\frac{1}{2}} \right\} \\ \nonumber
&& + \frac{1}{{C_{V_i}}^n_{i+\frac{1}{2},j+\frac{1}{2},k+\frac{1}{2}}} \Biggl[ \frac{{p_i}^n_{i+\frac{1}{2},j+\frac{1}{2},k+\frac{1}{2}} + q^n_{i+\frac{1}{2},j+\frac{1}{2},k+\frac{1}{2}}}{\rho^n_{i+\frac{1}{2},j+\frac{1}{2}+k\frac{1}{2}}}  \\ \nonumber
&& \left\{ \left( \frac{\partial u}{\partial x} \right)^n_{i+\frac{1}{2},j+\frac{1}{2},k+\frac{1}{2}} + \left( \frac{\partial v}{\partial y} \right)^n_{i+\frac{1}{2},j+\frac{1}{2},k+\frac{1}{2}}  + \left( \frac{\partial w}{\partial z} \right)^n_{i+\frac{1}{2},j+\frac{1}{2},k+\frac{1}{2}} \right\} \Biggr] \Biggr]
\end{eqnarray}

\begin{eqnarray*}
&& \left( u \frac{\partial {T_i}}{\partial x} \right)^n_{i+\frac{1}{2},j+\frac{1}{2},k+\frac{1}{2}} \\
&& \ \ \ \  = \left\{ 
\begin{array}{ll}
	u^n_{i+\frac{1}{2},j+\frac{1}{2},k+\frac{1}{2}} \frac{{T_i}^n_{i+\frac{1}{2},j+\frac{1}{2},k+\frac{1}{2}} -  {T_i}^n_{i-\frac{1}{2},j+\frac{1}{2},k+\frac{1}{2}}}{Dx^n_{i,j+\frac{1}{2},k+\frac{1}{2}}} &\left( u^n_{i+\frac{1}{2},j+\frac{1}{2},k+\frac{1}{2}} \geq 0 \right) \\
	u^n_{i+\frac{1}{2},j+\frac{1}{2},k+\frac{1}{2}} \frac{{T_i}^n_{i+\frac{3}{2},j+\frac{1}{2},k+\frac{1}{2}} - {T_i}^n_{i+\frac{1}{2},j+\frac{1}{2},k+\frac{1}{2}}}{Dx^n_{i+1, j+\frac{1}{2},k+\frac{1}{2}}} &\left( u^n_{i+\frac{1}{2},j+\frac{1}{2},k+\frac{1}{2}} <0 \right)
\end{array} 
\right. \\
&& \left( v \frac{\partial {T_i}}{\partial y} \right)^n_{i+\frac{1}{2},j+\frac{1}{2},k+\frac{1}{2}} \\
&& \ \ \ \  = \left\{ 
\begin{array}{ll}
	v^n_{i+\frac{1}{2},j+\frac{1}{2},k+\frac{1}{2}} \frac{{T_i}^n_{i+\frac{1}{2},j+\frac{1}{2},k+\frac{1}{2}} - {T_i}^n_{i+\frac{1}{2},j-\frac{1}{2},k+\frac{1}{2}}}{Dy^n_{i\frac{1}{2},j,k+\frac{1}{2}}} &\left( v^n_{i+\frac{1}{2},j+\frac{1}{2},k+\frac{1}{2}} \geq 0 \right) \\
	v^n_{i+\frac{1}{2},j+\frac{1}{2},k+\frac{1}{2}} \frac{{T_i}^n_{i+\frac{1}{2},j+\frac{3}{2},k+\frac{1}{2}} - {T_i}^n_{i+\frac{1}{2},j+\frac{1}{2},k+\frac{1}{2}}}{Dy^n_{i+\frac{1}{2}, j+1,k+\frac{1}{2}}} &\left( v^n_{i+\frac{1}{2},j+\frac{1}{2},k+\frac{1}{2}} <0 \right)
\end{array} 
\right. \\
&& \left( w \frac{\partial {T_i}}{\partial z} \right)^n_{i+\frac{1}{2},j+\frac{1}{2},k+\frac{1}{2}} \\
&& \ \ \ \  = \left\{ 
\begin{array}{ll}
	w^n_{i+\frac{1}{2},j+\frac{1}{2},k+\frac{1}{2}} \frac{{T_i}^n_{i+\frac{1}{2},j+\frac{1}{2},k+\frac{1}{2}} - {T_i}^n_{i+\frac{1}{2},j+\frac{1}{2},k-\frac{1}{2}}}{Dz^n_{i\frac{1}{2},j+\frac{1}{2},k}} &\left( w^n_{i+\frac{1}{2},j+\frac{1}{2},k+\frac{1}{2}} \geq 0 \right) \\
	w^n_{i+\frac{1}{2},j+\frac{1}{2},k+\frac{1}{2}} \frac{{T_i}^n_{i+\frac{1}{2},j+\frac{1}{2},k+\frac{3}{2}} - {T_i}^n_{i+\frac{1}{2},j+\frac{1}{2},k+\frac{1}{2}}}{Dz^n_{i+\frac{1}{2}, j+\frac{1}{2},k+1}} &\left( w^n_{i+\frac{1}{2},j+\frac{1}{2},k+\frac{1}{2}} <0 \right)
\end{array} 
\right. \\
&& \left( \frac{\partial u}{\partial x} \right)^n_{i+\frac{1}{2},j+\frac{1}{2},k+\frac{1}{2}} = \frac{u^n_{i+1,j+\frac{1}{2},k+\frac{1}{2}} - u^n_{i,j+\frac{1}{2},k+\frac{1}{2}}}{Dx^n_{i,j+\frac{1}{2},k+\frac{1}{2}}} \\
&& \left( \frac{\partial v}{\partial y} \right)^n_{i+\frac{1}{2},j+\frac{1}{2},k+\frac{1}{2}} = \frac{v^n_{i+\frac{1}{2},j+1,k+\frac{1}{2}} - v^n_{i+\frac{1}{2},j,k+\frac{1}{2}}}{Dy^n_{i+\frac{1}{2},j,k+\frac{1}{2}}} \\
&& \left( \frac{\partial w}{\partial z} \right)^n_{i+\frac{1}{2},j+\frac{1}{2},k+\frac{1}{2}} = \frac{w^n_{i+\frac{1}{2},j+\frac{1}{2},k+1} - w^n_{i+\frac{1}{2},j+\frac{1}{2},k}}{Dz^n_{i+\frac{1}{2},j+\frac{1}{2},k}}
\end{eqnarray*}
The electron temperature and radiation temperature are also expressed in the same way.
%\end{document} 	%Discretization
%\documentclass{jarticle}
%\begin{document}
	\section{Three temperature relaxation}
	In this study, we employ the three-temperature model for the ion, the electron and the radiation \cite{Tahir}. In this model, it assumed the radiation is in its equilibrium. The assumption means that the radiation becomes the Planck distribution. For example, the heavy ion beams (HIBs) deposit their energy inside a a material of the energy absorber \cite{kawata2, kawata3, kawata4}. The temperature in the energy absorber becomes around 300eV during the HIBs pulse length of $\sim$ ten ns, and all the three temeratures are almost equlibrated during the fusion fuel target implosion. However, at the fuel ignition and burning phases induced by the energy deposition of the alpha particles created by the DT fusion reactions, the three temperatures may be different among them. We need to compute the energy transfer between the three temperatures.  \par

	The following equations are used for the basic equation\cite{Tahir}. 
	\begin{eqnarray}
		\left\{ \begin{array}{lll}
			C_{V_i}\frac{dT_i}{dt}=-K_{ie}\\
			C_{V_e}\frac{dT_e}{dt}=K_{ie}-K_{re}\\
			C_{V_r}\frac{dT_r}{dt}=K_{re} \\
		\end{array} \right.
	\end{eqnarray}
	\\
	Here $C_{V_i}$ is ion constant volume specific heat$[{\rm J/K\cdot kg}]$. 
	$C_{V_e}$ is electron constant volume specific heat$[{\rm J/K\cdot kg}]$. 
	$C_{V_r}$ is radiation constant volume specific heat$[{\rm J/K\cdot kg}]$.
	$T_i$ is ion temprature$[{\rm K}]$, $T_e$ is electron temprature$[{\rm K}]$ and
	$T_r$ is radiation temprature$[{\rm K}]$.
	$K_{ie}$ is energy exchange rate between the ions and the electrons, and
	$K_{re}$ is energy exchange rate between the radiation and the electrons.\\
	
	The energy exchange rate is expressed by the following equation.
	\begin{eqnarray}
		\left\{ \begin{array}{ll}
			K_{ie}=C_{V_i}\omega_{ie}(T_i-T_e)\\
			K_{re}=C_{V_r}\omega_{re}(T_e-T_r)\\
		\end{array} \right.
	\end{eqnarray}

\noindent	
Here $\omega_{ie}$ and $\omega_{re}$ are the collision frequencies between the ion-electron and the radiation-electron, respectively. They are calculated by the following equations. The inverse Compton scattering is also included in the collision frequency between the radiation and the electrons.
	\begin{eqnarray}
		&&\omega_{ie}=\frac{Z^2e^4n\log{\Lambda}\sqrt{m_e}}{32\sqrt{2}\pi\varepsilon^2_0Mm_p(kT)^{3/2}}=6.57578\times10^{-10}\times\frac{n_i\log{\Lambda}Z^2}{MT^{3/2}_e}\ [\rm 1/s]\\ \nonumber \\
		&&\omega_{re}=\omega'_{re}+\omega_{cr}\nonumber\\
		&&\ \ \ \omega'_{re}=8.5\times10^{-14}\frac{\langle Z^2\rangle\langle Z\rangle n_iIg}{MT^{1/2}_ec_e}\ [{\rm 1/s}]\\
		&&\ \ \ \ \ Ig=\int^{\infty}_0\frac{\xi(e^{\xi u}-e^u)}{(\xi-1)(e^{\xi u}-1)(e^u-1)}du\nonumber \\
		&&\ \ \ \omega_cr=\frac{128}{3}\frac{\pi e^4\sigma}{(m_ec^2)^3}T^4_r=7.362\times10^{-22}T^4_r\ [{\rm 1/s}]
	\end{eqnarray}
	
\noindent	
Here, $u=\displaystyle\frac{h\nu}{kT_e},\ \xi=\displaystyle\frac{T_e}{T_i}$, $h$ is Planck's constant, and $\nu$ is the radiation frequency. 

	When the basic equations are discretized, they are expresses as follows: 
	\begin{eqnarray}
		\left\{ \begin{array}{lll}
			C^{n+\frac{1}{2}}_{V_i}\displaystyle\frac{T^{n+1}_i-T^*_i}{\Delta t^{n+\frac{1}{2}}}=-K^{n+\frac{1}{2}}_{ie}\\[10pt]
			C^{n+\frac{1}{2}}_{V_e}\displaystyle\frac{T^{n+1}_e-T^*_e}{\Delta t^{n+\frac{1}{2}}}=K^{n+\frac{1}{2}}_{ie}-K^{n+\frac{1}{2}}_{re}\\[10pt]
			C^{n+\frac{1}{2}}_{V_r}\displaystyle\frac{T^{n+1}_r-T^*_r}{\Delta t^{n+\frac{1}{2}}}=-K^{n+\frac{1}{2}}_{re}\\
		\end{array} \right.
	\end{eqnarray}

\noindent
Here $T^*$ indicates the temperatures after the calculation of the energy equations in Subsection \ref{Discretization of energy equation}. 

	By introducing the expressions of  $\xi_{ie}=T_i-T_e,\ \xi_{re}=T_e-T_r$, the energy exchange rates are expressed as follows: 
	\begin{eqnarray}
		\left\{ \begin{array}{ll}
			K^{n+\frac{1}{2}}_{ie}=C_{V_i}\omega^{n+\frac{1}{2}}_{ie}\xi^{n+\frac{1}{2}}_{ie}\\
			K^{n+\frac{1}{2}}_{re}=C_{V_r}\omega^{n+\frac{1}{2}}_{re}\xi^{n+\frac{1}{2}}_{re}\\
		\end{array} \right.
	\end{eqnarray}

\noindent
Here $\xi^{n+\frac{1}{2}}_{ie}$ and $\xi^{n+\frac{1}{2}}_{re}$ are expressed as follows: 
	\begin{eqnarray}
		&&\xi^{n+\frac{1}{2}}_{ie}=C_iA+\bigg[\xi^n_{ie}-\Bigl(\frac{\alpha_i}{\gamma}\Bigr)^{n+\frac{1}{2}}\bigg]B+\Bigl(\frac{\alpha_i}{\gamma}\Bigr)^{n+\frac{1}{2}} \\
		&&\xi^{n+\frac{1}{2}}_{re}=C_rA+\bigg[\xi^n_{re}-\Bigl(\frac{\alpha_r}{\gamma}\Bigr)^{n+\frac{1}{2}}\bigg]B+\Bigl(\frac{\alpha_r}{\gamma}\Bigr)^{n+\frac{1}{2}}
	\end{eqnarray}

\noindent
Here each symbol definition is displayed below: 
	\begin{eqnarray*}
		&&\alpha_i=(\phi_i+\beta_r\phi_r)\omega_{re}\\
		&&\alpha_r=(\phi_iG+\beta_i\phi_r)\omega_{ie}\\
		&&\beta_i=1+\frac{C_{V_i}}{C_{V_e}}\\
		&&\beta_r=1+\frac{C_{V_e}}{C_{V_r}}\\
		&&G=\frac{C_{V_i}}{C_{V_e}}\\
		&&\gamma=(\beta_i\beta_r-G)\omega_{ie}\omega_{re}\\
		&&A=\frac{\bigg[\exp{\Bigl(X\Delta t^{n+\frac{1}{2}}\Bigr)}-1\bigg]}{X\Delta t^{n+\frac{1}{2}}}-\frac{\bigg[\exp{\Bigl(Y\Delta t^{n+\frac{1}{2}}\Bigr)}-1\bigg]}{Y\Delta t^{n+\frac{1}{2}}}\\
		&&B=\frac{\bigg[\exp{\Bigl(Y\Delta t^{n+\frac{1}{2}}\Bigr)}-1\bigg]}{Y\Delta t^{n+\frac{1}{2}}}\\
		&&X=-\frac{1}{2}\lambda+\frac{1}{2}(\lambda^2-4\gamma)^{n+\frac{1}{2}}\\
		&&Y=-\frac{1}{2}\lambda-\frac{1}{2}(\lambda^2-4\gamma)^{n+\frac{1}{2}}\\
		&&\lambda=\beta_i\omega_{ie}+\beta_r\omega_{re}\\
		&&C_i=\frac{1}{(\lambda^2-4\gamma)^{n+\frac{1}{2}}}\bigg[\phi_i-\beta_i\omega_{ie}\xi_{i0}+\omega_{re}\xi_{r0}+\frac{1}{2}\lambda\Bigl(\xi_{i0}-\frac{\alpha_i}{\gamma}\Bigr)\bigg]+\frac{1}{2}\lambda\Bigl(\xi_{i0}-\frac{\alpha_i}{\gamma}\Bigr)\\
		&&C_r=\frac{1}{(\lambda^2-4\gamma)^{n+\frac{1}{2}}}\bigg[\phi_r-\beta_r\omega_{re}\xi_{r0}+G\omega_{ie}\xi_{i0}+\frac{1}{2}\lambda\Bigl(\xi_{r0}-\frac{\alpha_r}{\gamma}\Bigr)\bigg]+\frac{1}{2}\lambda\Bigl(\xi_{r0}-\frac{\alpha_r}{\gamma}\Bigr)\\
		&&\phi_i=\frac{W_i}{C_{V_i}}-\frac{W_e}{C_{V_e}}\\
		&&\phi_r=\frac{W_e}{C_{V_e}}-\frac{W_r}{C_{V_r}}\\
	\end{eqnarray*}

%	\end{document} 	%Three temperature relaxation
\newpage
%\documentclass{jarticle}
%\usepackage {amsmath}
%\usepackage{bm}
%\usepackage[left=30mm,right=30mm,top=35mm,bottom=35mm]{geometry}
%\usepackage{comment}
%\begin{document}
	\section{Heat conduction}
	The heat conduction is also solved to include the energy transport inside the target materials. \cite{Christiansen}.
	\begin{eqnarray}
		&&C_{V_k}\frac{DT}{Dt}=\frac{1}{\rho}{\bm \nabla}\cdot(\kappa_k{\bm \nabla}T_k)\ \ \ \ \ \ \ (k=i,e,r)\\
		&&\ \ \ \ \kappa_i=4.3\times10^{-12}T^{5/2}_{i}(\log{\Lambda})m^{-1/2}Z^{-4}\ \ [{\rm W/mK}]\\	\nonumber
		&&\ \ \ \ \kappa_e=1.83\times10^{-10}T^{5/2}_{e}(\log{\Lambda})^{-1}Z^{-1}\ \ [{\rm W/mK}]\\	\nonumber
		&&\ \ \ \ \kappa_r=\frac{16}{3}\sigma L_RT^3_r\ \ [{\rm W/mK}] \nonumber
	\end{eqnarray}
	The variables are defined as follows: 
		$\kappa_k$ is The thermal conductivity, 
		$T_k$ represents one of the temperatures for the ions, electrons and the radiation$[{\rm K}]$, 
		$\log{\Lambda}$ the Coulomb logarithm,  
		$m$ the atomic weight, 
		$Z$ the ionization degree.
		$\sigma$ the Stefan-Boltzmann constant, and 
		$L_R$ is the Rosseland mean free path \cite{Zeldovich}.
	\par
	The thermal conductivity  $k_r$ of the radiation is expressed together with a flux limit approximation. The energy flux should be limited to prevent an excess energy transport by a steep temperature gradient in ICF.  
	\begin{eqnarray}
		k_r = k_r (1 + \frac{4}{5} \frac{L_R}{T_r} \delta T_r )^{-1}
	\end{eqnarray}
	The basic equation is shown again below.
	\begin{eqnarray}
	\label{eq:heat1}
		C_V\frac{DT}{Dt}&=&\frac{1}{\rho}{\bm \nabla}\cdot(\kappa{\bm \nabla}T)
	\end{eqnarray}

	When Eq. (\ref{eq:heat1}) is discretized, the following equation is obtained.
	\begin{eqnarray}
		\frac{T^{n+1}_{i+\frac{1}{2},j+\frac{1}{2},k+\frac{1}{2}}-T^{n}_{i+\frac{1}{2},j+\frac{1}{2},k+\frac{1}{2}}}{dt^{n+\frac{1}{2}}} &=& \frac{1}{M^{n}_{i+\frac{1}{2},j+\frac{1}{2},k+\frac{1}{2}}{C_V}^{n}_{i+\frac{1}{2},j+\frac{1}{2},k+\frac{1}{2}}}	\nonumber\\
		&\times& \left\{ \left( \kappa^n_{i+1,j+\frac{1}{2},k+\frac{1}{2}}\frac{T^n_{i+\frac{3}{2},j+\frac{1}{2},k+\frac{1}{2}}-T^n_{i+\frac{1}{2},j+\frac{1}{2},k+\frac{1}{2}}}{Dx^2} \right. \right. \nonumber \\	
		 &-& \left. \kappa^n_{i,j+\frac{1}{2},k+\frac{1}{2}} \frac{T^n_{i+\frac{1}{2},j+\frac{1}{2},k+\frac{1}{2}}-T^n_{i-\frac{1}{2},j+\frac{1}{2},k+\frac{1}{2}}}{Dx^2}  \right) \nonumber \\
		 &+&  \left( \kappa^n_{i+\frac{1}{2},j+1,k+\frac{1}{2}}\frac{T^n_{i+\frac{1}{2},j+\frac{3}{2},k+\frac{1}{2}}-T^n_{i+\frac{1}{2},j+\frac{1}{2},k+\frac{1}{2}}}{Dy^2} \right. \nonumber \\
		 &-& \left. \kappa^n_{i+\frac{1}{2},j,k+\frac{1}{2}} \frac{T^n_{i+\frac{1}{2},j+\frac{1}{2},k+\frac{1}{2}}-T^n_{i+\frac{1}{2},j-\frac{1}{2},k+\frac{1}{2}}}{Dy^2}  \right)  \nonumber \\
		 &+&  \left( \kappa^n_{i+\frac{1}{2},j+\frac{1}{2},k+1}\frac{T^n_{i+\frac{1}{2},j+\frac{1}{2},k+\frac{3}{2}}-T^n_{i+\frac{1}{2},j+\frac{1}{2},k+\frac{1}{2}}}{Dz^2} \right. \nonumber  \\
		 &-& \left. \left. \kappa^n_{i+\frac{1}{2},j+\frac{1}{2},k} \frac{T^n_{i+\frac{1}{2},j+\frac{1}{2},k+\frac{1}{2}}-T^n_{i+\frac{1}{2},j+\frac{1}{2},k-\frac{1}{2}}}{Dz^2}  \right) \right\} 
	\end{eqnarray}
	
\begin{comment}
	\begin{eqnarray}
	\label{eq:CoDis}
	\begin{split}
		\frac{T^{n+1}_{k+\frac{1}{2},l+\frac{1}{2}}-T^{n}_{k+\frac{1}{2},l+\frac{1}{2}}}{dt^{n+\frac{1}{2}}}\!&=\!A_1\Bigl(T^{n+1}_{k+\frac{1}{2},l+\frac{1}{2}}\!-T^{n+1}_{k+\frac{1}{2},l-\frac{1}{2}}\Bigr)\!+\!A_2\Bigl(T^{n+1}_{k+\frac{1}{2},l+\frac{3}{2}}\!-T^{n+1}_{k+\frac{1}{2},l+\frac{1}{2}}\Bigr)\\
		&+\!B_1\Bigl(T^{n+1}_{k+\frac{1}{2},l+\frac{1}{2}}\!-T^{n+1}_{k-\frac{1}{2},l+\frac{1}{2}}\Bigr)\!+\!B_2\Bigl(T^{n+1}_{k+\frac{3}{2},l+\frac{1}{2}}\!-T^{n+1}_{k+\frac{1}{2},l+\frac{1}{2}}\Bigr)
	\end{split}
	\end{eqnarray}
	\\
	\\
ただし,
	\begin{eqnarray}
		&&A_1=-S^{n}_{k+\frac{1}{2},l}\bigg|\Delta\bar{\bm R}^{n}_{k+\frac{1}{2},l}\bigg|\frac{\kappa^{n}_{k+\frac{1}{2},l}}{j^{n}_{k+\frac{1}{2},l}}\ \ \ \ \ \ \ A_2=S^{n}_{k+\frac{1}{2},l+1}\bigg|\Delta\bar{\bm R}^{n}_{k+\frac{1}{2},l+1}\bigg|\frac{\kappa^{n}_{k+\frac{1}{2},l+1}}{j^{n}_{k+\frac{1}{2},l+1}}\\
		&&B_1=-S^{n}_{k,l+\frac{1}{2}}\bigg|\delta\bar{\bm R}^{n}_{k,l+\frac{1}{2}}\bigg|\frac{\kappa^{n}_{k,l+\frac{1}{2}}}{j^{n}_{k,l+\frac{1}{2}}}\ \ \ \ \ \ \ \ B_2=S^{n}_{k+1,l+\frac{1}{2}}\bigg|\delta\bar{\bm R}^{n}_{k+1,l+\frac{1}{2}}\bigg|\frac{\kappa^{n}_{k+1,l+\frac{1}{2}}}{j^{n}_{k+1,l+\frac{1}{2}}}\\
	\end{eqnarray}
\newpage
\end{comment}

The equations of the heat conduction is solved  by the ADI (Alternating Directional Implicit) method \cite{NumericalRecipes}. 
\begin{comment}
	\begin{align}
	\label{eq:ADIk}
		&\frac{T^{*}_{k+\frac{1}{2},l+\frac{1}{2}}-T^{n}_{k+\frac{1}{2},l+\frac{1}{2}}}{dt^{n+\frac{1}{2}}}\!=\!A_1\Bigl(T^{*}_{k+\frac{1}{2},l+\frac{1}{2}}-T^{n*}_{k+\frac{1}{2},l-\frac{1}{2}}\Bigr)\!+\!A_2\Bigl(T^{*}_{k+\frac{1}{2},l+\frac{3}{2}}-T^{*}_{k+\frac{1}{2},l+\frac{1}{2}}\Bigr)\\
	\label{eq:ADIl}
		&\frac{T^{n+1}_{k+\frac{1}{2},l+\frac{1}{2}}-T^{*}_{k+\frac{1}{2},l+\frac{1}{2}}}{dt^{n+\frac{1}{2}}}\!=\!B_1\Bigl(T^{n+1}_{k+\frac{1}{2},l+\frac{1}{2}}-T^{n+1}_{k-\frac{1}{2},l+\frac{1}{2}}\Bigr)\!+\!B_2\Bigl(T^{n+1}_{k+\frac{3}{2},l+\frac{1}{2}}\!-T^{n+1}_{k+\frac{1}{2},l+\frac{1}{2}}\Bigr)
	\end{align}
\end{comment}

%\end{document} 	%Heat conduction
%\documentclass[dvipdfmx]{jarticle}
%\usepackage{amsmath}
%\usepackage{bm}
%\usepackage[left=30mm,right=30mm,top=35mm,bottom=35mm]{geometry}

%\begin{document}
\section{Fusion reaction}
\subsection{Fusion reaction}
In the document we focus mainly on the reaction of the deuterium (D) and tritium (T). Additionally, the DD reaction is considered. The reaction equations are shown below.
	\begin{eqnarray}
	\begin{split}
        	\rm D+\rm T&\rightarrow{\rm He}^4(3.5{\rm MeV})+\rm n(14.1{\rm MeV})\\
		\rm D+\rm D&\xrightarrow[50\%]{}\rm T(1.01{\rm MeV})+\rm p(3.02{\rm MeV})\\
		&\xrightarrow[50\%]{}{\rm He}^3(0.82{\rm MeV})+\rm n(2.45{\rm MeV})
		\end{split}
		\label{NucEq}
	\end{eqnarray}

The number of reactions $N_{DT}$ per unit time in the D-T reaction is represented by the following equation from equation.
	\begin{equation}
	\label{eq:NucReacDT2}
		N_{\rm DT}=\langle\sigma v\rangle_{\rm DT}n_{\rm D}n_{\rm T}
	\end{equation}
	
	 Similarly, the number of reactions $N_{\rm DD}$ per unit time in the D-D reaction is expressed by the following equation.
	\begin{equation}
	\label{eq:NucReacDD2}
		N_{\rm DD}=\frac{1}{2}\langle\sigma v\rangle_{\rm DD}n_{\rm D}n_{\rm D}
	\end{equation}
	
	According to the formula (\ref{NucEq}), D is reduced by the D-D reaction and the D-T reaction. When Eqs. (\ref{eq:NucReacDT2})  and (\ref{eq:NucReacDD2}) are used, the amount of change in the number density $n_{\rm D}$ of D per minute time is expressed as follows.
	\begin{eqnarray}
	\label{eq:ReacD}
		\frac{\partial n_{\rm D}}{\partial t}&=&-N_{\rm DD}-N_{\rm DT}\nonumber\\
							 &=&-\frac{1}{2}\langle\sigma v\rangle_{\rm DD}n_{\rm D}n_{\rm D}-\langle\sigma v\rangle_{\rm DT}n_{\rm D}n_{\rm T}
	\end{eqnarray}
Further, according to Eq. (\ref{NucEq}), T is created by the DD reaction and consumed by the DT reaction.
Therefore, the number density $n_{\rm T}$ of T is expressed as follows: 
	\begin{eqnarray}
	\label{eq:ReacT}
		\frac{\partial n_{\rm T}}{\partial t}&=&+\frac{1}{2}N_{\rm DD}-N_{\rm DT}\nonumber\\
							&=&+\frac{1}{4}\langle\sigma v\rangle_{\rm DD}n_{\rm D}n_{\rm D}-\langle\sigma v\rangle_{\rm DT}n_{\rm D}n_{\rm T}
	\end{eqnarray}
	From Eq. (\ref{NucEq}) ${\rm He}^4$, that is, the $\alpha$ particle is generated by the DT reaction. 	\begin{eqnarray}
	\label{eq:alpha}
		\frac{\partial n_\alpha}{\partial t}&=&+N_{\rm DT}\nonumber\\
								&=&+\langle\sigma v\rangle_{\rm DT}n_{\rm D}n_{\rm T}
	\end{eqnarray}
	The $\alpha$ particles collide with the target ions and electrons during the diffusion process.
Then, the diffusion of the $\alpha$ particles and the $\ alpha$ particle energy deposition are expressed by the following equation: 
	\begin{equation}
	\label{eq:alphaReac}
		\frac{\partial n_\alpha}{\partial t}=+\langle\sigma v\rangle_{\rm DT}n_{\rm D}n_{\rm T}-\bm\nabla\cdot\bm F-\omega_\alpha n_\alpha
	\end{equation}
	
	\subsection{Reaction rate}
	Here, the reaction rates of the DD reaction and the DT reaction are described. In this study, the fusion reaction is calculated using the analytical curves corresponding to each reaction rate\cite{kawata4, Huba}. The formulae fitted are shown below: 
	\begin{eqnarray}
	\label{eq:AnalysisDD}
		&&\langle\sigma v\rangle_{\rm DD}=\exp{\left(x_1-\frac{x_2}{T_i^{x_5}}+\frac{x_3T_i}{(T_i+x_4)^2}\right)}\\
	\label{eq:AnalysisDT}
		&&\langle\sigma v\rangle_{\rm DT}=\exp{\left(x_1-\frac{x_2}{T_i^{x_5}}+\frac{x_3}{1+x_4}\right)}
	\end{eqnarray}
	Here $T_i$ is the ion temprature 
	
	The coefficients $x_n (n=1\sim5)$ in Eq. (\ref{eq:AnalysisDD}) are listed below for the 50\% of the DD reaction: 
	\begin{eqnarray*}
		x_1&=&-49.1789720673151\\[-6pt]
		x_2&=&15.3267580380585\\[-6pt]
		x_3&=&-4168271.58512757\\[-6pt]
		x_4&=&36677.9694366768\\[-6pt]
		x_5&=&0.365303247159742
	\end{eqnarray*}
	For another 50\% of the DD reactions, the coefficients $x_n (n=1\sim5)$ in Eq. (\ref{eq:AnalysisDD}) are listed below: 
	\begin{eqnarray*}
		x_1&=&-48.9931165228571\\[-6pt]
		x_2&=&15.6125104498645\\[-6pt]
		x_3&=&-4168271.58512757\\[-6pt]
		x_4&=&36677.9694366768\\[-6pt]
		x_5&=&0.363023326564475
	\end{eqnarray*}
	The corresponding coeficients for the D-T reaction in Eq. (\ref{eq:AnalysisDT}) are shown as follows: 
	\begin{eqnarray*}
		x_1&=&-48.9580509680824\\[-6pt]
		x_2&=&18.1155080330636\\[-6pt]
		x_3&=&895.149425658926\\[-6pt]
		x_4&=&135.888636700177\\[-6pt]
		x_5&=&0.366290140624939
	\end{eqnarray*}
%	\end{document}		%Fusion reaction
\newpage
%\documentclass[dvipdfmx]{jarticle}
%\usepackage{pdfpages}
%\usepackage{bm}
%\begin{document}

\section{$\alpha$ particle heating}
\subsection{$\alpha$ particle diffusion}
The flux of the $\alpha$ particle is shown below $\mbox{\boldmath $F$}$.
\begin{eqnarray}
\mbox{\boldmath $F$}  = - D_\alpha \mbox{\boldmath $\nabla$} n_\alpha
\end{eqnarray}
Here $D_\alpha$ is the diffusion coefficient and is expressed by the following equation.
\begin{eqnarray}
\label{eq:alpha_kakusan}
D_\alpha = \frac{ \frac{1}{3} v_\alpha \lambda_\alpha}{1+ \frac{4}{3} \lambda_\alpha \frac{|\nabla n_\alpha|}{n_\alpha}}
\end{eqnarray}
Here $v_\alpha$ is the speed of $\alpha$ particle and $\lambda_\alpha$ the mean free path of $\alpha$． The second term of the denominator in Eq. (\ref{eq:alpha_kakusan}) expresses the flux limiting effect, which limits the excess flux by the steep gradient of the $\alpha$ density. The flux $\mbox{\boldmath $F$}$ of the $\alpha$ particles in the $x$，$y$ and $z$ directions are expressed by the following equations:  
\begin{eqnarray}
F_x = - \frac{ \frac{1}{3} n_\alpha v_\alpha \lambda_\alpha }{n_\alpha + \frac{4}{3} \lambda_\alpha \left| \frac{\partial n_\alpha}{\partial x} \right| } \frac{\partial n_\alpha}{\partial x} \\
F_y = - \frac{ \frac{1}{3} n_\alpha v_\alpha \lambda_\alpha }{n_\alpha + \frac{4}{3} \lambda_\alpha \left| \frac{\partial n_\alpha}{\partial y} \right| } \frac{\partial n_\alpha}{\partial y} \\
F_z = - \frac{ \frac{1}{3} n_\alpha v_\alpha \lambda_\alpha }{n_\alpha + \frac{4}{3} \lambda_\alpha \left| \frac{\partial n_\alpha}{\partial z} \right| } \frac{\partial n_\alpha}{\partial z} 
\end{eqnarray}

\subsection{$\alpha$ particle deposition}
　In the fusion target plasma, the $\alpha$ particles collide with the ions and the electrons. When only the collision term is considered, it is expressed by the following equation: 
\begin{eqnarray}
\frac{\partial n_\alpha}{\partial t} = -n_\alpha \omega_\alpha
\end{eqnarray}
During the short time interval of $dt$, we can assume that $\omega_\alpha$ is constant. Then an analytical solution is obtained. 

\begin{eqnarray}
n_\alpha \propto e^{-\omega_\alpha t}
\end{eqnarray}

The energy deposited to the electrons and the ions are expressed by the following equation: 
	\begin{equation}
		\rho C_v\Delta T=+E_\alpha n_\alpha f
	\label{AplphaDepositionEq}
	\end{equation}
			
	In Eq. (\ref{AplphaDepositionEq}), $f$ represents the partition ratio between the ions and the electrons. Here $f_i$ is the $\alpha$-particle energy deposition ratio to the  ions, and $f_e$ the deposition ratio to the electrons \cite{Fraley}.
		\begin{eqnarray}
		f_i&=&\frac{1}{1+\displaystyle\frac{32}{T_e}}\\
		f_e&=&1-f_i
	\end{eqnarray}

The energy increase by the $\alpha$ particle energy deposition is expressed by the following equation: 
		\begin{eqnarray}
	\label{eq:NucEnIon}
		\Delta T_i=\frac{E_\alpha n_\alpha f_i}{\rho C_{v_i}}\\
	\label{eq:NucEnEle}
		\Delta T_e=\frac{E_\alpha n_\alpha f_e}{\rho C_{v_e}}
	\end{eqnarray}
%\end{document}		%alpha particle
%\documentclass[dvipdfmx]{jarticle}
%\begin{document}

\section{Algorithm review}
Here we summarize the computation cycle.
\begin{enumerate}
\item Initial setup and computation preparations. 
\item The time step $dt$ is controlled to avoid the numerical instability. $dt = C_{FL}min\{dl/(V+C_s)\}$. The time step of $dt$ is evaluated at each mesh direction. The minimum $dt$ is employed. Here $V$ shows the absolute value of the plasma velocity and $C_s$ the sound speed. In plasmas and fluids shock waves may appear, and the shock speed would be larger than the sound speed. The coefficient of $C_{FL}$ should be less than 1.0. Normally we use $C_{FL} < 0.1$ to ensure the numerical stability and the numerical accuracy. 
\item The artificial viscosity $q^{n+1}_{i+\frac{1}{2},j+\frac{1}{2},k+\frac{1}{2}}$ is obtained.
\item The velocity $u^{n+1}_{i,j+\frac{1}{2},k+\frac{1}{2}}$, $v^{n+1}_{i+\frac{1}{2},j,k+\frac{1}{2}}$ and $w^{n+1}_{i+\frac{1}{2},j+\frac{1}{2},k}$ are obtained by equation of motion.
\item The mass density $\rho^{n+1}_{i+\frac{1}{2},j+\frac{1}{2},k+\frac{1}{2}}$ is obtained by equation of continuity.
\item The temperature $T^{n+1}_{i+\frac{1}{2},j+\frac{1}{2},k+\frac{1}{2}}$ is obtained by energy equation.
\item The pressure $p^{n+1}_{i+\frac{1}{2},j+\frac{1}{2},k+\frac{1}{2}}$, the specific heat $C^{n+1}_{Vi+\frac{1}{2},j+\frac{1}{2},k+\frac{1}{2}}$ and the compressibility $B^{n+1}_{Ti+\frac{1}{2},j+\frac{1}{2},k+\frac{1}{2}}$ are obtained by equation of state.
\item The $\alpha$ particles are created by the DT fusion reactions, and the $\alpha$ particles are diffused. The new temperature $T^{n+1}_{i+\frac{1}{2},j+\frac{1}{2},k+\frac{1}{2}}$ is obtained by the $\alpha$ particle energy deposition.
\item The new temperature $T^{n+1}_{i+\frac{1}{2},j+\frac{1}{2},k+\frac{1}{2}}$ is obtained by the temperature relaxation among the three temperatures. In addition, the electron energy and radiation energy are conducted. 

\end{enumerate}
%\end{document}
\section{Summary}
In this document we described a numerical algorithm for a 3D Euler fluid code with a uniform spatial mesh to simulate the nuclear fusion fuel ignition and burning. At the fusion fuel stagnation, ignition and burning phases, the fusion fuel spatial deformation would be serious. In order to avoid the spatial mesh crush, we use the Euler method in these phases, though the implosion phase can be simulated by the Lagrange method \cite{kawata4, kawata5} or the ALE (Arbitrary Lagrangian Eulerlian) method \cite{Milan}. 

\renewcommand{\refname}{References}

\end{document}